\documentclass[prd,nofootinbib]{revtex4-2}

\usepackage{amsmath, amssymb, amsthm, graphicx, epsfig, fancyhdr,epsfig, slashed}

\usepackage{tikzsymbols}
\usepackage{natbib}
\usepackage{float}

\usepackage{tikz,xcolor,hyperref}

\definecolor{lime}{HTML}{A6CE39}
\DeclareRobustCommand{\orcidicon}{
	\begin{tikzpicture}
	\draw[lime, fill=lime] (0,0) 
	circle [radius=0.2] 
	node[white] {{\fontfamily{qag}\selectfont \tiny ID}};
	\draw[white, fill=white] (-0.0625,0.095) 
	circle [radius=0.007];
	\end{tikzpicture}
	\hspace{-2mm}
}

\foreach \x in {A, ..., Z}{\expandafter\xdef\csname orcid\x\endcsname{\noexpand\href{https://orcid.org/\csname orcidauthor\x\endcsname}
			{\noexpand\orcidicon}}
}

\newcommand{\be}{\begin{equation}}
\newcommand{\ee}{\end{equation}}
\newcommand{\bea}{\begin{eqnarray}}
\newcommand{\eea}{\end{eqnarray}}

\newcommand{\ag}[1]{\textcolor{blue}{[Anish: #1]}}

\begin{document}


\title{Non-perturbative Origin of the Electroweak Scale with Dyson-Schwinger: \\ \it{Nambu-Jona-Lasino-like Fermionic Mass Gap and Higher-order Excitations}}

\author{Marco Frasca}
\email{marcofrasca@mclink.it}
\affiliation{Rome, Italy}

\author{Anish Ghoshal}
\email{anish.ghoshal@roma2.infn.it}
\affiliation{Institute of Theoretical Physics, Faculty of Physics, \\ University of Warsaw, ul. Pasteura 5, 02-093 Warsaw, Poland}

 \author{Nobuchika Okada}
 \email{okadan@ua.edu}
 \affiliation{Department of Physics and Astronomy, \\ University of Alabama, Tuscaloosa, AL 35487, USA}

\begin{abstract}
\textit{We study the Yukawa interaction of a fermion field $\psi$ with a scalar field $\phi$ and analyze the spectrum of the theory. It is shown that due to non-perturbative dynamics, $\phi$ develops a vacuum expectation value (vev) in the form of a mass gap which can potentially trigger the electroweak symmetry breaking (EWSB) and dynamically generates the SM Higgs boson mass. For estimating the non-perturbatively generated mass scale, we solve the hierarchy of Dyson-Schwinger Equations in the form of partial differential equations using the exact solution known via a novel technique developed by Bender, Milton and Savage. We employ Jacobi Elliptic function as exact background solution and show that the mass gap that arises in the fermion sector (with Nambu-Jona-Lasino-like framework) can be transmuted to the EW sector, expressed in terms of fermion mass and the self-quartic coupling of $\phi$. }
\end{abstract}

\maketitle

\section{Introduction}

With no evidence for supersymmetric or any new particles at the TeV energy scales in LHC, there maybe alternative mechanism at work that may hint towards resolution or alleviating the Higgs naturalness problem. One such alternative direction of model building is to promote scale invariance to be a  symmetry of the fundamental classical action and be only broken at the quantum level paving way for dynamical generation of the electroweak (EW) scale and the SM Higgs mass. In weakly coupled theories this may happen via the famous Coleman-Weinberg mechanism or radiative symmetry breaking \cite{Coleman:1973jx,Holthausen:2009uc,Meissner:2006zh,Foot:2007as,Farzinnia:2013pga}. On the other hand, in a strongly interacting sector leading to non-perturbative physics this maybe possible via what is known as dimensional transmutation and have been widely studied involving BSM hidden gauge groups for the strongly-interacting dark sector \cite{Holthausen:2013ota,Hambye:2007vf,Hur:2011sv,Hambye:2013dgv}\footnote{For mass generation via extended scalar sector see Ref. \cite{Romatschke:2024hpb}}. For instance the standard quantum chromodynamics (QCD)  e.g.~\cite{Kubo:2014ova} along with additional matter content charged under QCD may do this job. Such a novel way of breaking EW symmetry by the condensation of chiral fermions in high color representation with new fermionic condensates are studied since the 1980s \cite{Marciano:1980zf,Zoupanos:1983xh,Lust:1985aw,Lust:1985jk} but currently in severe tension with the latest tests of EW precision observables at LHC and beyond \cite{Peskin:1990zt,Zyla:2020zbs}. 

We suggest in this paper a modified scenario hidden sector fermions which are a singlet under the EW gauge group extension. These fermions are not chiral (for instance vector-like) so that its mass is already technically natural as we know that the chiral symmetry protects the fermionic mass term \cite{tHooft:1979rat}. In this paper a novel non-perturbative tool to treat such strongly interacting fermionic sector. We study the condensate and dynamical chiral symmetry breaking (DCSB) in the framework of  Dyson-Schwinger equations (DSEs) of strongly-coupled interactions. Our considerations outlined in this paper are based on the exact solution of
the background equations of motion in Yang-Mills theory in terms of Jacobi elliptic functions 
following the analytic approach of Dyson-Schwinger
equations, originally devised by Bender, Milton and
Savage~\cite{Bender:1999ek}. We are able to represent the Green's functions of the theory analytically, therefore we understand the effect of the background on the interactions that remains valid even in the strongly-coupled regime~\cite{Frasca:2015yva}. This tool has already been widely applied to QCD~\cite{Frasca:2021yuu,Frasca:2021mhi,Frasca:2022lwp,%
Frasca:2022pjf,Chaichian:2018cyv} and to the scalar
sector~\cite{Frasca:2015wva}. Moreoever extensions beyond the traditional QFT and SM of particle physics to other types of
models involving new gauge sectors and string-inspired non-local
theories have been widely discussed using this methodology~\cite{Frasca:2019ysi,Chaichian:2018cyv,Frasca:2017slg,Frasca:2016sky,Frasca:2015yva,Frasca:2015wva,Frasca:2013tma,Frasca:2012ne,Frasca:2009bc,Frasca:2010ce,Frasca:2008tg,Frasca:2009yp,Frasca:2008zp,Frasca:2007uz,Frasca:2006yx,Frasca:2005sx,Frasca:2005mv,Frasca:2005fs}. As a practical application and predictions related to experiments concerning
particle physics phenomenology and cosmology, like for instance, non-perturbative hadronic vacuum polarization contributions to the muon anomalous magnetic moment $(g-2)_\mu$ \cite{Frasca:2021yuu}, QCD in the non-perturbative regimes~\cite{Frasca:2021mhi,Frasca:2022lwp,Frasca:2022pjf}, Higgs-Yukawa theory \cite{Frasca:2023qii}, finite temperature field theory \cite{Frasca:2023eoj},
non-perturbative false vacuum decay and phase transitions~\cite{Frasca:2022kfy,Calcagni:2022tls,Calcagni:2022gac}, dark energy \cite{Frasca:2022vvp}, and investigations
of the mass gap and confinement in string-inspired infinite-derivative and Lee-Wick theories~\cite{Frasca:2020jbe,Frasca:2020ojd,Frasca:2021iip, Frasca:2022duz,Frasca:2022gdz} have been considered extensively and proved to be quite successful. These tools were also employed to implement dynamical generation Hiigs mass via scalar and vector condensates too very recently \cite{Frasca:2024fuq,Chatterjee:2024dgw,Frasca:2024pmv,Frasca:2023eoj}.

The paper is organized as follows: in section 2, we discuss the model, in section 3 we compute the partition function of the strongly coupled theory, in section 4 we show the generation of mass gap considering a Nambu-Jona-Lasino like scenario and finally we end with conclusion and discussion.

\medskip


\section{Model and Lagrangian}

We consider the following Lagrangian, with a real scalar $\phi$,
\be
L=-\frac{1}{2}\phi\Box \phi+V(\phi)+{\bar\psi}(i{\slashed\partial}-g \phi-m_0)\psi.
\ee
Our aim is to get the fermionic spectrum of the theory. This can be worked out by noting that the quantum theory of the $\phi$ field has been extensively discussed in \cite{Frasca:2015yva}. Indeed, for the sake of simplicity we are assuming $V(\phi)=\lambda(\phi^4/4$ with the understanding that the $\phi$ field could get a mass by self-interaction and interaction through the fermionic field. Besides, the analysis is limited to a single component of the scalar sector.

We want to solve the quantum equation of motion
\begin{equation}
    \partial^2\phi+\lambda\phi^3=j
\end{equation}
with given the partition function
\begin{equation}
    Z[j]=\int[d\phi]e^{-\int d^4x\left[\frac{1}{2}(\partial\phi)^2-\frac{\lambda}{4}\phi^4+j\phi\right]}.
\end{equation}
From this, the correlation functions can be computed by the following set of Dyson-Schwinger equations \cite{Frasca:2015yva}.
The 1-point function is obtained by
\begin{equation}
\label{eq:g10}
   \partial^2 G_1(x)+\lambda\left([G_1(x)]^3+3G_2(x,x)G_1(x)+G_3(x,x,x)\right)=0.
\end{equation}
For the 2-point function one has
\begin{eqnarray}
\label{eq:g20}
   &&\partial^2G_2(x,y)+\lambda\left(3[G_1(x)]^2G_2(x,y)+3G_2(x,x)G_2(x,y)\right. \nonumber \\
	&&\left.+3G_3(x,x,y)G_1(x)+G_4(x,x,x,y)\right)=\delta^4(x-y).
\end{eqnarray}
The computation of the partition function entails the possibility to get a Gaussian solution. By this, we mean that all the higher-order correlation function with $n>2$ are completely determined by the 1P- and 2P-correlation functions that are solutions of the above equations. This can only be achieved if we assume $G_n(x_1,x_2,\ldots,x_n)=0$ if and only if they are evaluated in at least two identical coordinates corresponding to a Gaussian partition function. We will see how this can happen by a property of the support of the $G_2$ distributions under integration in a discussion below for the $G_3$ and $G_4$ correlation functions. Given this assumption, Eq.(\ref{eq:g10}) admits two kind of solutions. For $G_1(x)=\text{constant}$ (Higgs solution)\footnote{This happens because, only after dimensional regularization, one can have $G_2(x,x)<0$ (see \cite{afgg2024}).}, translation invariance is preserved. Otherwise, we get the following solution, a Fubini-Lipatov instanton solution, that breaks such a symmetry
\begin{equation}
\label{eq:G1}
    G_1(x)=\sqrt{\frac{2\mu^4}{m^2+\sqrt{m^4+2\lambda\mu^4}}}{\rm sn}\left(p\cdot x+\chi,\frac{-m^2+\sqrt{m^4+2\lambda\mu^4}}{-m^2-\sqrt{m^4+2\lambda\mu^4}}\right)
\end{equation}
with $\mu$ and $\chi$ being arbitrary integration constants, $m^2=3\lambda G_2(x,x)$, $G_3(x,x,x)=0$ and the momenta $p$ 
\begin{equation}
\label{eq:disp}
    p^2=m^2+\frac{\lambda\mu^4}{m^2+\sqrt{m^4+2\lambda\mu^4}}.
\end{equation}
It is important to notice that the dispersion relation would be massive even if setting  $G_2(x,x)=0$ it would be possible. 
Given this solution, it is very easy to obtain the two-point function in momenta space \cite{Frasca:2013tma} and recover translation invariance. If we neglect the mass shift due to renormalization in a first instance, we get \cite{Frasca:2015yva}
\begin{equation}
\label{eq:G2}
   G_2(p)=\frac{\pi^3}{4K^3(-1)}
	\sum_{n=0}^\infty\frac{e^{-(n+\frac{1}{2})\pi}}{1+e^{-(2n+1)\pi}}(2n+1)^2\frac{1}{p^2-m_n^2+i\epsilon}
\end{equation}
and the mass spectrum
\begin{equation}
\label{eq:ms}
   m_n=(2n+1)\frac{\pi}{2K(-1)}\left(\frac{\lambda}{2}\right)^\frac{1}{4}\mu.
\end{equation}
This will solve the equation for $G_2$ provided we, consistently, will have $G_3(x,x,y)=0$ and $G_4(x,x,x,y)=0$ in the following. Indeed, we have, after currents are set to zero,
\begin{eqnarray}
    &&\partial^2G_3(x,y,z)+\lambda\left[6G_1(x)G_2(x,y)G_2(x,z)+3G_1^2(x)G_3(x,y,z)\right. \\ \nonumber
    &&+3G_2(x,z)G_3(x,x,y)+3G_2(x,y)G_3(x,x,z) \\ \nonumber
    &&\left.+3G_2(x,x)G_3(x,y,z)+3G_1(x)G_4(x,x,y,z)+G_5(x,x,x,y,z)\right]=0 \\ \nonumber
    &&\\ \nonumber
    &&\partial^2G_4(x,y,z,w)+\lambda\left[6G_2(x,y)G_2(x,z)G_2(x,w)
    \right. \\ \nonumber
    &&+6G_1(x)G_2(x,y)G_3(x,z,w)+6G_1(x)G_2(x,z)G_3(x,y,w)\\ \nonumber
    &&+6G_1(x)G_2(x,w)G_3(x,y,z)+3G_1^2(x)G_4(x,y,z,w) \\ \nonumber
    &&+3G_2(x,y)G_4(x,x,z,w)+3G_2(x,z)G_4(x,x,y,w)  \\ \nonumber
    &&+3G_2(x,w)G_4(x,x,y,z)+3G_2(x,x)G_4(x,y,z,w) \\ \nonumber
    &&\left.+3G_1(x)G_5(x,x,y,z,w)+G_6(x,x,x,y,z,w)\right]=0 \\ \nonumber
    &\vdots&
\end{eqnarray}
that are solved by
\begin{equation}
\label{eq:G_3}
   G_3(x,y,z)=-6\lambda\int dx_1 G_2(x,x_1)G_1(x_1,y)G_2(x_1,y)G_2(x_1,z) ,
\end{equation}
and it is easy to verify that $G_3(x,x,z)=G_3(x,y,x)=0$ using the property of Heaviside function $\theta(x)\theta(-x)=0$, and
\begin{eqnarray}
    &&G_4(x,y,z,w)=-6\lambda\int dx_1 G_2(x,x_1)G_2(x_1,y)G_2(x_1,z)G_2(x_1,w) \\ \nonumber
    &&-6\lambda\int dx_1G_2(x,x_1)\left[G_1(x_1,y)G_2(x_1,y)G_3(x_1,z,w)\right. \\ \nonumber
    &&\left.+G_1(x_1)G_2(x_1,z)G_3(x_1,y,w)
    +G_1(x_1,y)G_2(x_1,w)G_3(x_1,y,z)\right].
\end{eqnarray}
Similarly, it is not difficult to verify that $G_4(x,x,x,y)=0$. This implies that we can expand the action in the partition function using
\bea
\label{eq:DSs}
\phi(x)&=&G_1(x)+\int d^4x_1G_2(x,x_1)j(x_1)+\frac{1}{2!}\int d^4x_1d^4x_2G_3(x,x_1,x_2)j(x_1)j(x_2)
\nonumber \\
&&+\frac{1}{3!}\int d^4x_1d^4x_2d^4x_3G_4(x,x_1,x_2,x_3)j(x_1)j(x_2)j(x_3)+O(j^4).
\eea
For our aims, we can stop at the second term.

\section{Partition Function}
\label{SecPF}

Using eq.(\ref{eq:DSs}), one has
\bea
Z[j,{\bar\eta},\eta]&=&\exp\left[-S_c[j]-\frac{1}{2}\int d^4x_1d^4x_2j(x_1)G_2(x_1-x_2)j(x_2)\right]
\times \\
&&\int [d{\bar\psi}][d\psi]
\exp\left[-\int d^4x{\bar\psi}(x)\left(-{\slashed\partial}-g^2\int d^4x_1G_2(x-x_1){\bar\psi(x_1)\psi(x_1)}\right)\psi(x)\right]\times \nonumber \\
&&\exp\left[-\int d^4x\left({\bar\eta}(x)\psi(x)+{\bar\psi}(x)\eta(x)\right)\right]+O(j^3).
\nonumber
\eea
After a Fierz rearrangement, for the fermionic part we get
\bea
{\bar\psi}(x){\bar\psi(x_1)\psi(x_1)}\psi(x)&=&
\frac{1}{4}\left({\bar\psi}(x)\psi(x){\bar\psi(x_1)\psi(x_1)}+\right. \nonumber \\
&&{\bar\psi}(x)\gamma_\mu\psi(x){\bar\psi(x_1)\gamma^\mu\psi(x_1)}-
\frac{1}{8}{\bar\psi}(x)[\gamma^\mu,\gamma^\nu]\psi(x){\bar\psi(x_1)[\gamma_\mu,\gamma_\nu]\psi(x_1)}-
\nonumber \\
&&\left.{\bar\psi}(x)\gamma^5\psi(x){\bar\psi(x_1)\gamma^5\psi(x_1)}+
{\bar\psi}(x)\gamma^5\gamma^\mu\psi(x){\bar\psi(x_1)\gamma^5\gamma_\mu\psi(x_1)} .
\right)
\eea
We see that, from a single fermionic field, we get five possible excitations out. For the moment, we limit our interest to the scalar-axial part and perform a Stratanovich-Hubbard transformation on the fermionic part as \cite{Ebert:1997fc}
\bea
\exp\left[\frac{g^2}{4}\int d^4x\int d^4x_1\left({\bar\psi}(x)\psi(x)G_2(x-x_1){\bar\psi(x_1)\psi(x_1)}-
{\bar\psi}(x)\gamma^5\psi(x)G_2(x-x_1){\bar\psi(x_1)\gamma^5\psi(x_1)}\right)\right]
&=& \nonumber \\
\int[d\sigma][d\pi]
\exp\left[-\frac{2}{g^2}\int d^4x\int d^4x_1\left(\sigma(x)G_2(x-x_1)\sigma(x_1)-
\pi(x)G_2(x-x_1)\pi(x_1)\right)\right.&& \nonumber \\
\left.-g\int d^4x{\bar\psi}(x)(\sigma(x)+\gamma^5\pi(x))\psi(x)\right].&&
\eea


The local limit can be obtained by taking $G_2(x_2-x_1) \propto \delta^4(x_2-x_1)$. Indeed, from Eq.~(\ref{eq:G2}) we take the limit $p\rightarrow 0$, and obtain
\be
\label{eq:G2L}
   G_{2L}(x-x_1)=-\frac{\pi}{K(-1)}\frac{1}{\mu^2}\sqrt{\frac{2}{\lambda}}
	\sum_{n=0}^\infty\frac{e^{-(n+\frac{1}{2})\pi}}{1+e^{-(2n+1)\pi}}\delta^4(x-x_1)
 =-G\delta^4(x-x_1).
\ee
Finally, we can integrate out the fermionic degrees of freedom to get
\be
L_{eff}=-\frac{1}{2G}(\sigma^2+\pi^2)-i\operatorname{tr}\ln({i\slashed\partial-m_f-g\sigma-g\gamma^5\pi}).
\ee

The fermionic determinant can be expanded as we notice that
\be
\operatorname{tr}\ln({i\slashed\partial-m_f-g\sigma-g\gamma^5\pi})=
\operatorname{tr}\ln({i\slashed\partial-m_f})+
\operatorname{tr}\ln(1-(i\slashed\partial-m_f)^{-1}(g\sigma+g\gamma^5\pi)).
\ee
We can neglect the first term (a constant) and expand the second logarithm to obtain
\be
\operatorname{tr}\ln(1-(i\slashed\partial-m_f)^{-1}(\sigma+\gamma^5\pi))=
\operatorname{tr}\left(
-(i\slashed\partial-m_f)^{-1}(g\sigma+g\gamma^5\pi)
-\frac{1}{2}(i\slashed\partial-m_f)^{-2}(g^2\sigma^2+g^2\pi^2+2g^2\gamma^5\sigma\pi)+\ldots
\right).
\ee
This will yield \cite{Ebert:1982pk}.
\bea
\label{eq:Leff}
L&=&-\frac{1}{2G}(\sigma^2+\pi^2)+\frac{1}{2}(8g^2I_1)(\sigma^2+\pi^2)+\frac{1}{2}(4g^2I_2)[(\partial\sigma)^2+(\partial\pi)^2] \\
&&-\frac{1}{2}4m_f^2(4g^2I_2)\sigma^2
-8g^3(8m_fI_2)\sigma(\sigma^2+\pi^2)-2g^4I_2(\sigma^2+\pi^2)^2+\ldots.
\nonumber
\eea
with
\bea
I_1(m)1&=&i\int^\Lambda\frac{d^4p}{(2\pi)^4}\frac{1}{p^2-m^2} \nonumber \\
I_2(m)&=&-i\int^\Lambda\frac{d^4p}{(2\pi)^4}\frac{1}{(p^2-m^2)^2}.
\eea
These are divergent integrals to be regularized by a cut-off $\Lambda$. 
We evaluate the above potential for $\sigma=m$ and $\pi=0$, so that
\be
\left.\frac{\delta L_{eff}}{\delta\sigma}\right|_{\sigma=m,\pi=0}=0.
\ee
This will yield immediately the gap equation \cite{Ebert:1997fc}
\be
m=m_f+8mGI_1(m).
\ee
This is the fermion mass. Therefore, in eq.(\ref{eq:Leff}) the first two terms cancel. This holds in the chiral limit where the initial mass of the fermion is 0. This also grants the $m_\pi=0$. So, we have a pseudoscalar field, a Goldstone boson, that is massless unless the fermion has a mass. Then, we introduce the renormalized fields
\be
\sigma\rightarrow (4g^2I_2)^\frac{1}{2}\sigma, \qquad \pi\rightarrow (4g^2I_2)^\frac{1}{2}\pi,
\ee
and we get the final result, after chiral symmetry breaking,
\bea
\label{eq:Leff2}
L&=&\frac{1}{2}[(\partial\sigma)^2+(\partial\pi)^2]-\frac{1}{2}(4m^2+m_f^2)\sigma^2 -\frac{1}{2}\frac{m_fm}{NI_2G}\pi^2\\
&&
-\frac{m}{I_2^\frac{1}{2}}\sigma(\sigma^2+\pi^2)-\frac{1}{8I_2}(\sigma^2+\pi^2)^2+\ldots.
\nonumber
\eea
This model is renormalizable.

\section{Mass gap and scalar self-coupling}

\subsection{Local case}

\begin{figure}[H]
\centering
\includegraphics[height=8cm,width=8cm]{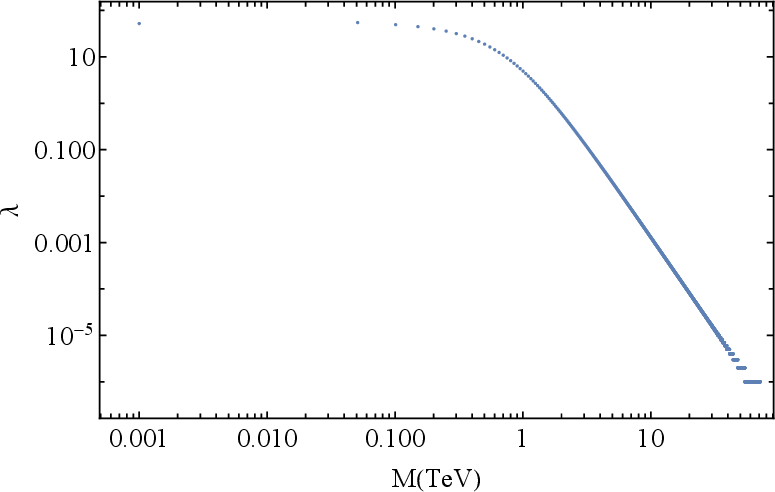}
\caption{\it The scalar self-coupling $\lambda$ as a function of fermion mass in the local case. To fix the ideas, we have chosen M to be around TeV scale. It is seen that larger is the scalar coupling and smaller is the fermion mass. We see a plateau at small values of the fermion mass. This could be an unphysical region where the gap equation has no solution for the fermion mass.}
\label{fig:MvsL}
\end{figure}

For the local case, we can get an implicit function for the fermion mass and the scalar self-coupling. One has \cite{Klevansky:1992qe}
\be
\label{eq:mgl}
M = m_f+\frac{1}{2\pi^2}\frac{NN_fM}{m_0^2+\frac{1}{G}}\left[\Lambda^2-M^2\ln\left(1+\frac{\Lambda^2}{M^2}\right)\right].
\ee
Indeed, from eq.(\ref{eq:G2L}) we get
\be
G=\kappa\frac{g^2}{\mu^2(\lambda/2)^\frac{1}{2}},
\ee
with $\kappa$ a proportionality constant of the order of unity, and
\be
\label{eq:m0}
m_0=\frac{\pi}{2K(i)}\left(\frac{\lambda}{2}\right)^\frac{1}{4}\mu.
\ee
This yields
\be
    \sqrt{\lambda} = \frac{M}{\sqrt{2}\pi^2(M-m_f)}\frac{NN_f}{\frac{\pi^2}{4K^2(i)}+\frac{1}{\kappa g^2}}\left[\frac{\Lambda^2}{\mu^2}-\frac{M^2}{\mu^2}\ln\left(1+\frac{\Lambda^2}{M^2}\right)\right]
\ee
The plot is given in fig.\ref{fig:MvsL}.

We see that larger is the self-coupling and smaller is the fermion mass unitl an unphysical region is reached.

\subsection{Non-local case}

In this case, we use the gluon propagator from eq.(\ref{eq:G2}) instead. 
Then, the gap equation can be written down in the form \cite{Frasca:2011bd}
\be
\label{eq:mg1}
M=m_f+{\cal C}(p)v
\ee
and
\be
\label{eq:mg2}
v=\frac{4N}{m_0^2+\frac{1}{G}}\int\frac{d^4p}{(2\pi)^4}{\cal C}(p)\frac{M(p)}{p^2+M^2(p)},
\ee
for a single flavor and $m_0$ given by eq.(\ref{eq:ms}). In our case the kernel is given by
\begin{equation}
   {\cal G}(p)=-\frac{1}{2}g^2\sum_{n=0}^\infty\frac{B_n}{p^2-(2n+1)^2(\pi/2K(i))^2\sigma+i\epsilon}=\frac{G}{2}{\cal C}(p).
\end{equation}
with $G$ being the Nambu-Jona-Lasinio constant that in our case is given by $G=2{\cal G}(0)=(g^2/m_0^2)\sum_{n=0}^\infty\frac{B_n}{(2n+1)^2(\pi/2K(i))^2}\propto (g^2/m_0^2)$, so that ${\cal C}(0)=1$. Here $m_0$ is obtained from eq.(\ref{eq:m0}).
The coefficients in the series are given by
\begin{equation}
   B_n=(2n+1)^2\frac{\pi^3}{4K^3(i)}\frac{e^{-(n+\frac{1}{2})\pi}}{1+e^{-(2n+1)\pi}}
\end{equation}
being $K(i)$ the complete elliptic integral of the first kind.

\begin{figure}[H]
\centering
\includegraphics[height=8cm,width=8cm]{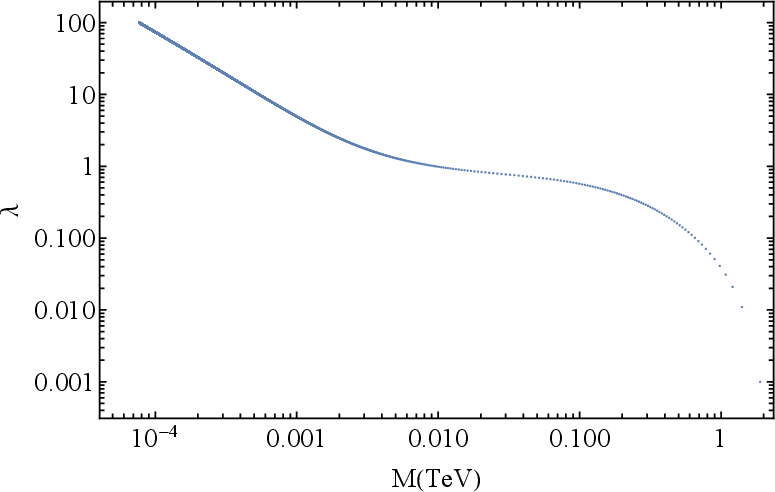}
\caption{\it The scalar self-coupling $\lambda$ as a function of the fermion mass  in the non-local case. We observe a similar behavior as for the local case. The plateau appears in a limited range of fermion masses. This is rather an inflection point.}
\label{fig:MvsnL}
\end{figure}

We have to solve eqs.(\ref{eq:mg1}) and (\ref{eq:mg2}) numerically and evaluate the fermion mass as a function of $\lambda$, the self-coupling of the Yukawa sector. $\lambda$ enters into the constant $\sigma$ as already said. IN fig.\ref{fig:MvsnL}, we show $\lambda$ as a function of fermion mass.
We observe a similar behavior as for the local case but the unphysical region disappears leaving just an inflection point.

\medskip


\section{Conclusions and Discussions}
\label{conc}

In order to understand the generation of the EWBS scale, we have considered a scalar field interaction with a fermion, commonly known as the Higgs-Yukawa model. We work with exact solutions of the scalar sector obtained by solving the corresponding set of Dyson-Schwinger equations through a partial differential equation (PDE) form according to Bender et. al. \cite{Bender:1999ek}. We summarise the salient features of our results :
\begin{enumerate}
    \item We get the spectrum of the theory taking the local limit of the scalar field propagator, that we know in closed form. We gend up with a Nambu-Jona-Lasinio model with the coupling constant completely given by the parameters of the Lagrangian we started from. This basically means that we have a pion-like field and a self-interacting scalar field generally termed as the $\sigma$-model in the literature and responsible for the breaking of the chiral symmetry, as discussed in Sec.~\ref{SecPF}.
    \item We show in this way that the scalar field gets a mass spectrum with a gap as the ground state is massive (see eq.(\ref{eq:m0})). Besides, a mass correction arising from quantum effects is also present as seen from eq.(\ref{eq:G1}).
    \item From the exact solution of the scalar sector, we are able to show that the fermion gets a mass through a gap equation that obtains solutions both in the local (see eq.(\ref{eq:mgl})) and in the non-local case (see eq.(\ref{eq:mg1}) and (\ref{eq:mg2})).
    \item Plots for the local and non-local cases show that the fermion mass becomes smaller as the self-coupling of the scalar field increases but it enters into an unphysical region for the local case. (see Fig.\ref{fig:MvsL} and \ref{fig:MvsnL}).
\end{enumerate}

We conclude that a Higgs-Yukawa model could be in principle effectively lead to dynamical generation of the EW symmetry breaking scale due to the mass generated in the fermion sector that interact with the Higgs. We leave a generalization to the standard model with all the realistic particle spectrum for future studies.


\section{Acknowledgement}
\label{Asck}
The work of NO is supported in part by the United States Department of Energy (DC-SC 0012447 and DC-SC 0023713).

\end{document}